\documentclass[a4paper,11pt]{article}
\usepackage{jinstpub}
\usepackage{lineno}
\usepackage{siunitx}
\usepackage{multirow}
\ifdefined\qty\else
  \ifdefined\NewCommandCopy
    \NewCommandCopy\qty\SI
  \else
    \NewDocumentCommand\qty{O{}mm}{\SI[#1]{#2}{#3}}
  \fi
\fi
\ifdefined\unit\else
  \ifdefined\NewCommandCopy
    \NewCommandCopy\unit\si
  \else
    \NewDocumentCommand\unit{O{}m}{\si[#1]{#2}}
  \fi
\fi

\title{\boldmath Lab and Beam Tests of a SiPM-readout ASIC with a Large Dynamic Range}

\author[a,b]{Baohua Qi,}
\author[a,b,1]{Yong Liu,\note{Corresponding author.}}
\author[c]{Huangchao Shi,}
\author[c]{Danqi Wang,}
\author[d,e,f]{Zhiyu Zhao}
\author[c]{and Hongbo Zhu}

\affiliation[a]{Institute of High Energy Physics, Chinese Academy of Sciences,\\19B Yuquan Road, 100049 Beijing, China}
\affiliation[b]{University of Chinese Academy of Sciences,\\19A Yuquan Road, 100049 Beijing, China}
\affiliation[c]{School of Physics, Zhejiang University,\\866 Yuhangtang Road, 310058 Hangzhou, China}
\affiliation[d]{Tsung-Dao Lee Institute, Shanghai Jiao Tong University,\\1 Lisuo Road, 201210 Shanghai, China}
\affiliation[e]{School of Physics and Astronomy, Shanghai Jiao Tong University,\\800 Dongchuan Road, 200240 Shanghai, China}
\affiliation[f]{Key Laboratory for Particle Astrophysics and Cosmology (MoE), Shanghai Key Laboratory for Particle Physics and Cosmology,\\800 Dongchuan Road, 200240 Shanghai, China}

\emailAdd{liuyong@ihep.ac.cn}

\abstract{A front-end readout system with a large dynamic range is required for the high-granularity crystal electromagnetic calorimeter (ECAL) at future Higgs factories. A new commercially available ASIC, MPT2321, which offers a significantly large dynamic range for the readout of the silicon photomultiplier (SiPM), has been tested in the laboratory and at a test beam facility. The fundamental performance metrics have been characterised, including response linearity, inter-calibration factor and noise performance. To quantitatively assess the dynamic range of the chip and evaluate its feasibility for SiPM readout in high-granularity crystal ECAL, a beamtest was conducted using scintillating crystals exposed to electron beams. Beamtest results showed that the chip exhibits a promising performance in terms of a good signal-to-noise ratio and a relatively large dynamic range.}


\arxivnumber{2411.18927}

\begin{document}
\maketitle
\flushbottom

\section{Introduction}
High-granularity calorimetry dedicated to future lepton colliders is expected to significantly enhance jet reconstruction performance by utilising 5D information (3D spatial, time, and energy). A novel particle-flow-oriented high-granularity crystal electromagnetic calorimeter (ECAL) has been proposed for experiments at future Higgs factories, e.g. the Circular Electron Positron Collider (CEPC)~\cite{cite:CEPCCDRVol2}. The crystal ECAL features a finely segmented structure and a silicon photomultiplier (SiPM) readout scheme, delivering an excellent electromagnetic performance, as demonstrated in recent studies~\cite{cite:CrystalECALCHEF2019,cite:CrystalECALCALOR2022}. Its homogeneous design enables high sensitivity to low-energy photons and neutral pions, while the unique high-granularity geometry aligns with the pattern recognition requirements of the particle-flow algorithm~\cite{cite:PFA}.

The readout electronics system for the crystal ECAL must address two stringent requirements: a high signal-to-noise ratio (SNR) to fully leverage the advantages of SiPMs, and a wide dynamic range to accommodate the energy deposition at \qty{360}{\giga \electronvolt} $t\bar{t}$ mode of the CEPC. The requirements are driven by the homogeneous calorimeter design, where individual crystals may receive substantial energy deposition and generate tens of thousands of photons, posing significant challenges to SiPMs and readout electronics. Specifically, each sensitive unit of the crystal ECAL must achieve a dynamic range spanning from 1 to \num{3e5} photon equivalent (\unit{\mathrm{p.e.}}), while ensuring a signal-to-noise ratio (SNR) greater than 3~\cite{cite:CrystalECALCALOR2022}. Commercially available chips designed for the SiPM readout, such as Petiroc~\cite{cite:Petiroc} and Citiroc~\cite{cite:Citiroc} developed by Omega laboratory and Weeroc, typically offer a dynamic range of up to \qty{3000}{\mathrm{p.e.}} (\qty{480}{\pico \coulomb}) with a typical SiPM gain of \num{1e6}~\cite{cite:PetirocCitirocTests}. Even with reduced SiPM gains on the order of \num{e5}, there remains a significant gap between the achievable dynamic range and the detector requirements. Thus, the front-end electronics system used for the readout of millions of SiPM channels has become one of the major instrumentation bottlenecks.

The novel high-precision Application-Specific Integrated Circuit (ASIC) MPT2321~\cite{cite:MicroParityMPT2321} developed by MicroParity is a promising candidate for the front-end electronics of the crystal ECAL. This chip features 32 readout channels designed specifically for SiPM-based applications, employing a 12-bit Analogue-to-Digital Converter (ADC), and a 20-bit Time-to-Digital Converter (TDC). The automatic high-low gain switching function ensures the sensitivity for measuring low-energy signals while accommodating a large dynamic range. The main specifications of the chip are presented in Table~\ref{tab:MPT2321Parameters}.

\begin{table}[tbp]
\centering
\caption{Main specifications of the MPT2321 chip.\label{tab:MPT2321Parameters}}
\smallskip
\begin{tabular}{l|l}
\hline
Number of channels & 32\\
ADC precision & \qty{1}{\milli \volt}\\
TDC precision & \qty{50}{\pico \second}\\
Signal shaping time & \qty{50}{\nano \second} or \qty{100}{\nano \second}\\
Maximum signal rate on single channel & $f_{\mathrm{SYSCLK}} / 60$\\
Data transmission standard & LVDS\\
Control interface & I$^2$C\\
Power supply & \qty{1.8}{\volt} \& \qty{3.3}{\volt}\\
Power consumption &  < \qty{500}{\milli \watt}\\
\hline
\end{tabular}
\end{table}

In this context, a series of characterisation experiments were conducted in the lab using an MPT2321 evaluation board, as presented in Section~\ref{sec:LabTests}, regarding the chip's response linearity, inter-calibration factor and noise performance. Section~\ref{sec:BeamTests} details the conduction of a beamtest using high-light-yield crystals and high-energy electrons, focusing on SiPM readout tests and dynamic range measurements for the chip. Finally, the summary and outlook are followed in Section~\ref{sec:Summary}.

\section{Lab Characterisations}
\label{sec:LabTests}
\begin{figure}[tbp]
\centering
\includegraphics[width=.49\textwidth]{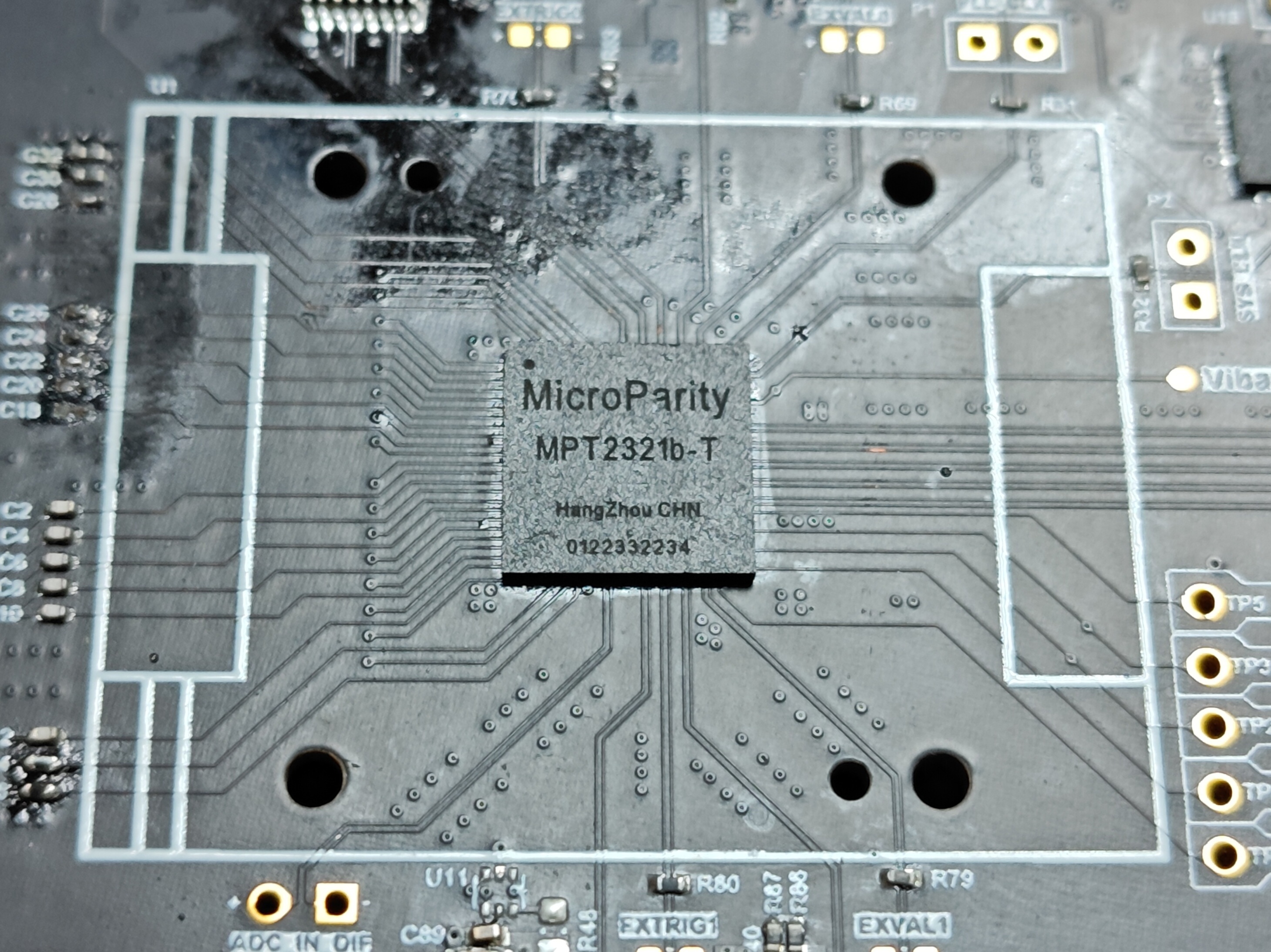}
\includegraphics[width=.49\textwidth]{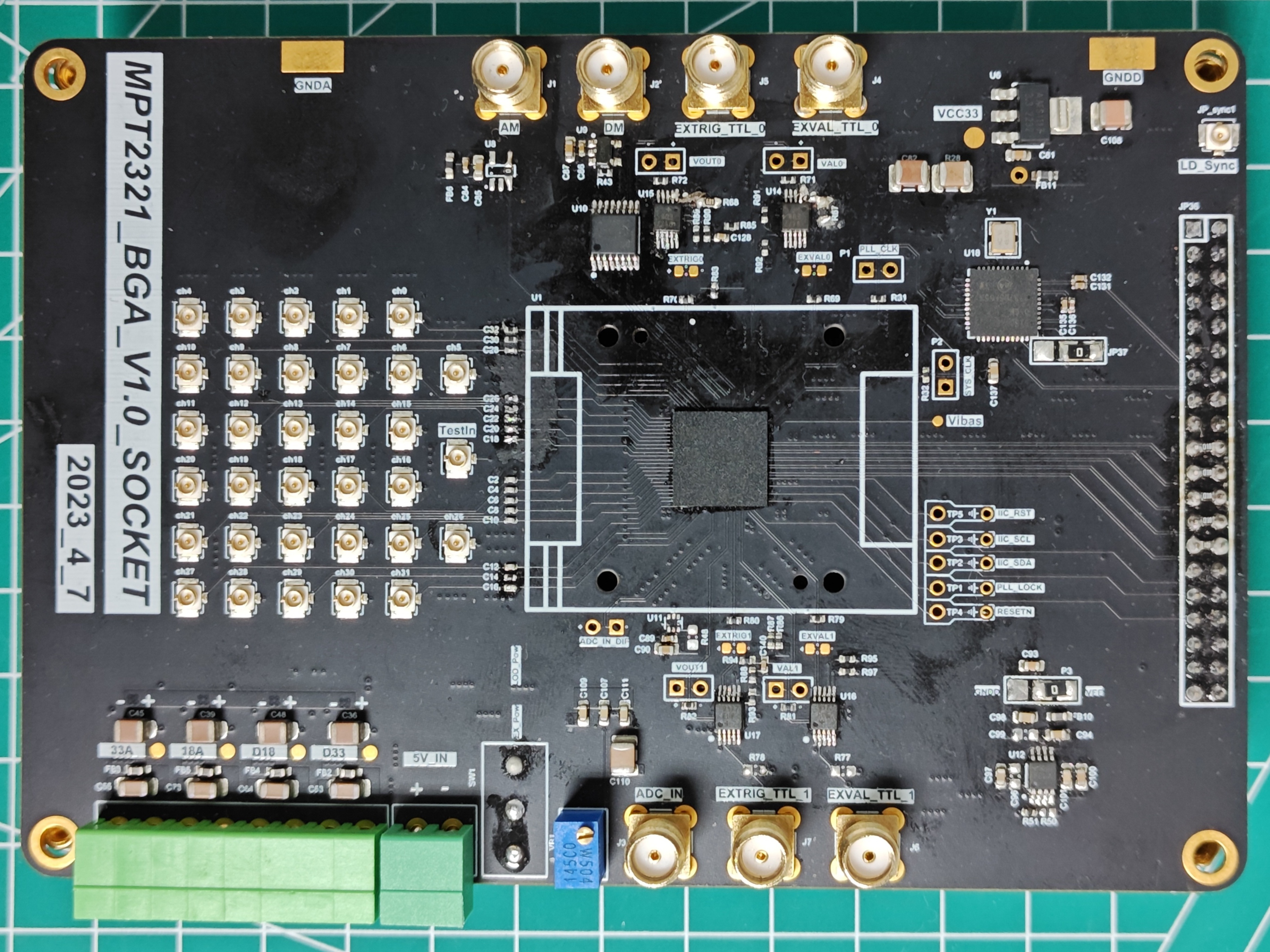}
\caption{The MPT2321 ASIC (left) and the evaluation board (right) for performance tests.\label{fig:MPT2321}}
\end{figure}

Figure~\ref{fig:MPT2321} shows an evaluation board equipped with an MPT2321 ASIC that has been utilised for performance characterisation. The chip incorporates 8 distinct gain modes, including 4 high-gain (HG) and 4 low-gain (LG) configurations. The high-gain modes are optimised to detect small signals and provide an excellent signal-to-noise ratio for the single photon calibrations. Meanwhile, the low-gain modes are designed to extend the dynamic range, allowing for effective measurements of SiPM signals with high scintillation light intensity. Automatic high-low gain switching was disabled to conduct a comprehensive characterisation for each gain mode individually. Signal shaping time has been set to \qty{100}{\nano \second} in the measurements.

\subsection{Response Linearity}
\label{sec:ResponseLinearity}
\begin{figure}[tbp]
\centering
\includegraphics[width=.98\textwidth]{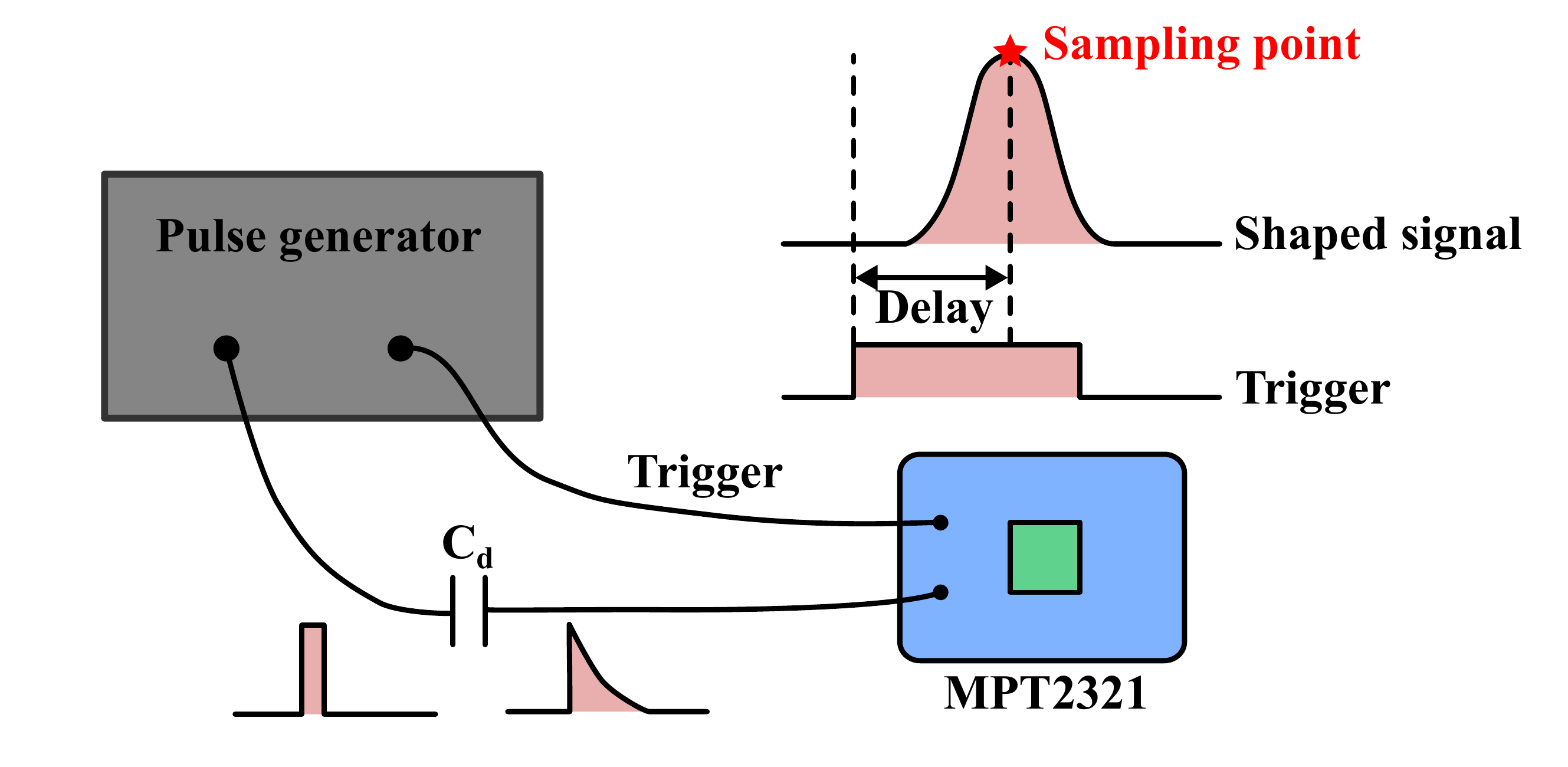}
\caption{Schematic diagram of the experimental setup for the charge injection tests.\label{fig:Exp_MPTChargeInjection}}
\end{figure}

The response linearity of the ASIC has been evaluated in the lab using the charge injection method. Figure~\ref{fig:Exp_MPTChargeInjection} shows the schematic view of the experimental setup. Square waves generated by a pulse generator are applied to a capacitor, producing a series of pulse signals with charge values determined by the product of capacitance and voltage. The signals are input into the chip for amplification and shaping. The synchronisation signal from the pulse generator serves as a trigger for the chip, and following a pre-calibrated delay time, the ADC samples the waveform at its peak to determine the corresponding charge value of the signal.

The capacitance of $\mathrm{C_d}$ for pulse waveform generation was determined according to the measurement requirements. For HG characterisation, a \qty{100}{\pico \farad} capacitor was selected based on the typical SiPM capacitance. Given the significant difference in gain factors between HG and LG modes, the LG configurations necessitated a \qty{1}{\nano \farad} capacitor to maintain sufficient voltage headroom for full dynamic range coverage, while generating waveforms that do not substantially diverge from the SiPM signals. The corresponding input voltages for charge injection tests are detailed in Table~\ref{tab:ChargeInjectionDetails}.

It should be noted that using different capacitances significantly affects the charge injection circuit's characteristics. Under the same signal acquisition and processing parameters (e.g. shaping time and delay time), the chip demonstrates different responses to circuits with varying characteristics, leading to inconsistent behaviour toward signals with the same charge amount. Experiments with \qty{100}{\pico \farad} and \qty{1}{\nano \farad} capacitors demonstrated that charge injection tests using different capacitances yield varying gain factors and inter-calibration factors. As a result, the measured response curves with the charge injection method are not suitable for determining calibration constants for systems utilising different components or configurations. Nonetheless, the measured response linearity remains valid, with no significant impact observed from these capacitance-related effects.

\begin{table}[tbp]
\centering
\caption{Experimental parameters of the charge injection test.\label{tab:ChargeInjectionDetails}}
\smallskip
\begin{tabular}{c|c|c|c}
\hline
\textbf{Configuration} & \textbf{$\mathrm{C_d}$} & \textbf{Input Voltage [\si{\milli \volt}]} & \textbf{Injected Charge [\si{\pico \coulomb}]}\\
\hline
HG Mode 1 & \multirow{4}{*}{\qty{100}{\pico \farad}} & \numrange{2}{70} & \numrange{0.2}{7}\\
HG Mode 2 & & \numrange{2}{75} & \numrange{0.2}{7.5}\\
HG Mode 3 & & \numrange{10}{240} & \numrange{1}{24}\\
HG Mode 4 & & \numrange{10}{480} & \numrange{1}{48}\\
\hline
LG Mode 1 & \multirow{4}{*}{\qty{1}{\nano \farad}} & \numrange{30}{540} & \numrange{30}{540}\\
LG Mode 2 & & \numrange{50}{1300} & \numrange{50}{1300}\\
LG Mode 3 & & \numrange{100}{2100} & \numrange{100}{2100}\\
LG Mode 4 & & \numrange{100}{4000} & \numrange{100}{4000}\\
\hline
\end{tabular}
\end{table}

\begin{figure}[tbp]
\centering
\includegraphics[width=.48\textwidth]{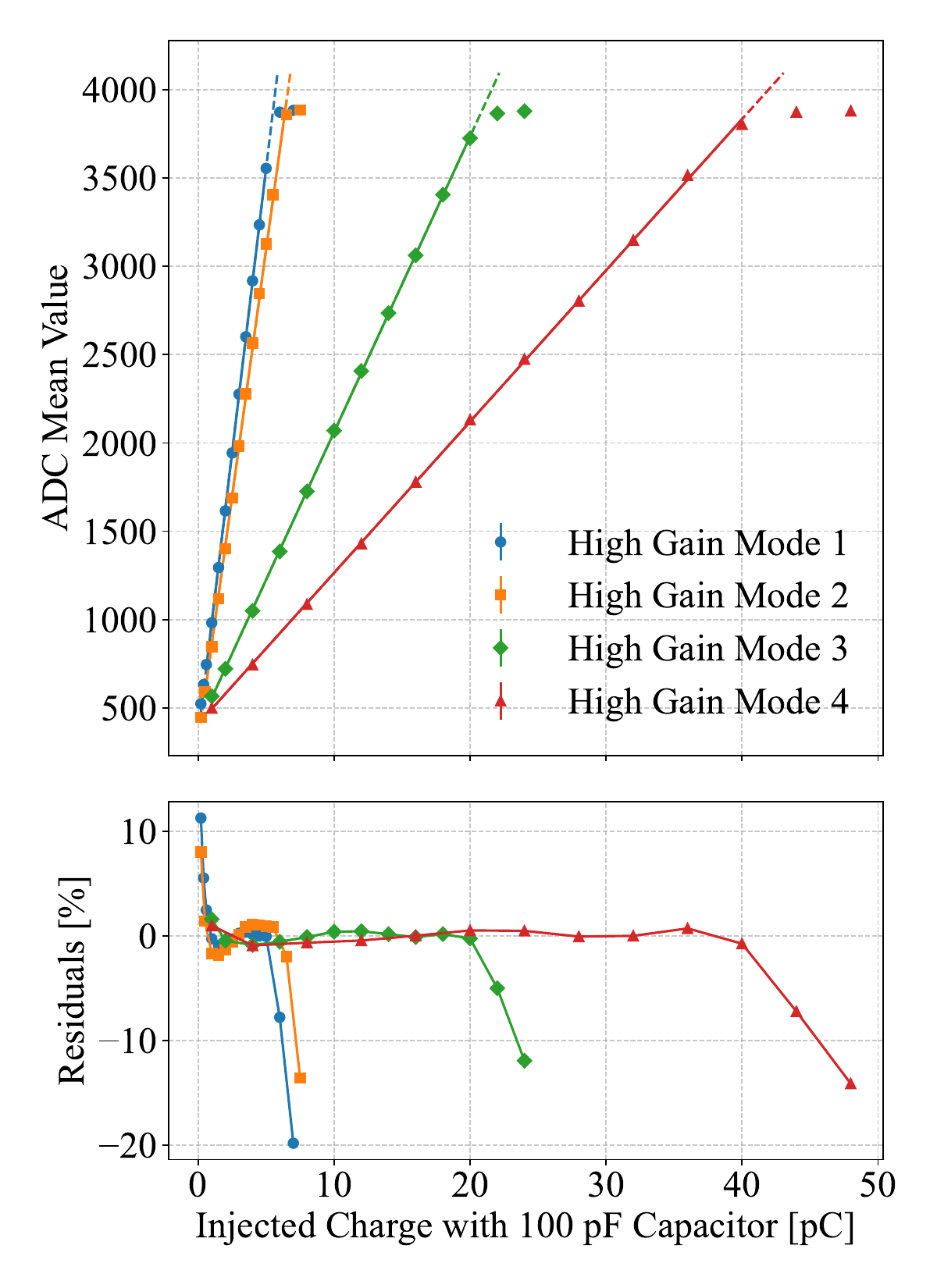}
\includegraphics[width=.48\textwidth]{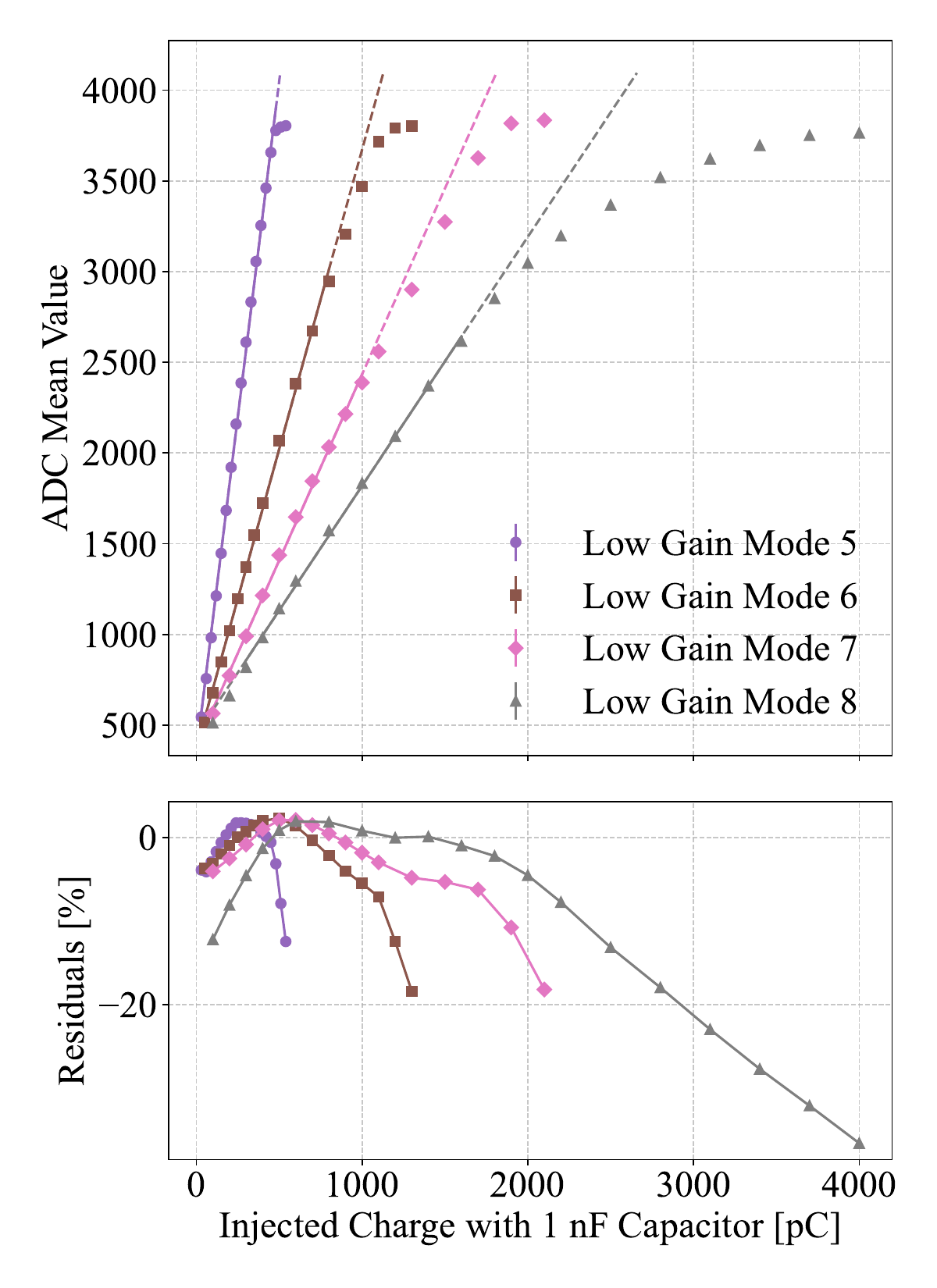}
\caption{Response linearity and residual curves under high-gain (left) and low-gain (right) modes. Charges were injected using \qty{100}{\pico \farad} and \qty{1}{\nano \farad} capacitors for high-gain and low-gain modes, respectively.\label{fig:ChargeInjection}}
\end{figure}

Figure~\ref{fig:ChargeInjection} demonstrates the ADC response to injected charge, including integral non-linearity (INL) defined as percentage deviation from linear fit. The linear response ranges are determined through several linear fitting iterations, with acceptance criteria defined by maximum deviation limits of $\pm2.5\%$ for HG modes and $\pm5\%$ for LG modes. This self-defined constraint is generally reasonable, as it aims to ensure that there are sufficient data points involved in the fitting while excluding regions exhibiting significant non-linearity. The HG modes exhibit significantly better linearity than LG modes, with full saturation consistently occurring at \qty{3800}{\mathrm{ADC}} counts. Non-linearity becomes apparent when a large number of charges are injected, mainly due to the ADC saturation effect. Meanwhile, persistent deviations at lower charge levels may result from the intrinsic ADC non-linearity and the contribution of noise. The slope parameters derived from linear fitting represent the gain factors corresponding to specific capacitance values. A comprehensive summary of these performance characteristics is provided in Table~\ref{tab:LinearResponseResults}.

The experimental setup, however, faces inherent limitations rooted in two critical parameters: the time constant (10-\qty{100}{\nano\second} for typical SiPMs) and the waveform shape (superpositions of avalanche signals). Such discrepancies become particularly evident when compared to crystal-coupled detector systems, where the crystal-induced photon arrival time spread dominates. Thus, a reliable performance evaluation requires direct characterisation using crystal-SiPM units under specific operational conditions.

Overall, the charge injection tests successfully determined the chip's linear range, and revealed critical characteristics of the ASIC: (a) HG modes demonstrate superior linearity ($\pm2.5\%$ INL), suitable for precision measurements; (b) LG modes achieve extended dynamic range for large signal measurements with controlled non-linearity ($\pm5\%$ INL); (c) All gain modes exhibit consistent saturation at approximately \qty{3800}{\mathrm{ADC}} counts. The observed capacitance-dependent gain factors necessitate system-specific calibration procedures for different detector configurations.

\begin{table}[tbp]
\centering
\caption{Linear performance characteristics of the MPT2321 chip obtained from charge injection tests.\label{tab:LinearResponseResults}}
\smallskip
\begin{tabular}{c|c|c|c|c}
\hline
\textbf{Configuration} & \textbf{Linear Range [\unit{\mathrm{ADC}}]} & \textbf{Injected Charge [\si{\pico \coulomb}]} & \textbf{Gain} & \textbf{Max. INL}\\
\hline
HG Mode 1 & \numrange{746}{3554} & \numrange{0.6}{5} & 642.95 & 2.46\%\\
HG Mode 2 & \numrange{590}{3857} & \numrange{0.5}{6.5} & 559.06 & 1.99\%\\
HG Mode 3 & \numrange{566}{3725} & \numrange{1}{20} & 167.25 & 1.57\%\\
HG Mode 4 & \numrange{500}{3804} & \numrange{1}{40} & 85.60 & 0.97\%\\
\hline
LG Mode 1 & \numrange{545}{3777} & \numrange{30}{480} & 7.41 & 4.11\%\\
LG Mode 2 & \numrange{517}{2945} & \numrange{50}{800} & 3.30 & 3.73\%\\
LG Mode 3 & \numrange{565}{2388} & \numrange{100}{1000} & 2.05 & 4.06\%\\
LG Mode 4 & \numrange{819}{2615} & \numrange{300}{1600} & 1.37 & 4.57\%\\
\hline
\end{tabular}
\end{table}

\subsection{Gain Inter-calibration}
\label{sec:GainInterCali}
As pointed out in Section~\ref{sec:ResponseLinearity}, the gain factors derived from charge injection tests exhibit capacitance-dependent characteristics, implying that the inter-calibration factors between different gain modes are not universally applicable. To obtain accurate inter-calibration factors tailored to a specific SiPM operational condition, a dedicated test was conducted using a SiPM and a laser source.

The test system comprised a \qty{405}{\nano\meter} picosecond laser (\qty{1}{\kilo \hertz} repetition rate) and a Hamamatsu S13360-3025PE SiPM~\cite{cite:HPKS13360} with a \qtyproduct{3x3}{\milli \meter \squared} active area, \qty{25}{\micro \meter} pixel pitch, and \qty{320}{\pico \farad} capacitance, operated at a \qty{5}{\volt} over-voltage. The ultrashort laser pulses ensured synchronous photon arrival at the SiPM, minimising timing-related variations. The laser was focused onto the SiPM, and the signal was directly output to the chip. Laser intensity was carefully controlled within the chip’s linear response range, and three distinct intensity levels were used to calibrate the high-gain, low-gain, and cross-gain relationships. Inter-calibration factors were determined by comparing signal amplitudes across gain configurations under identical laser intensities.

\begin{figure}[tbp]
\centering
\includegraphics[width=.48\textwidth]{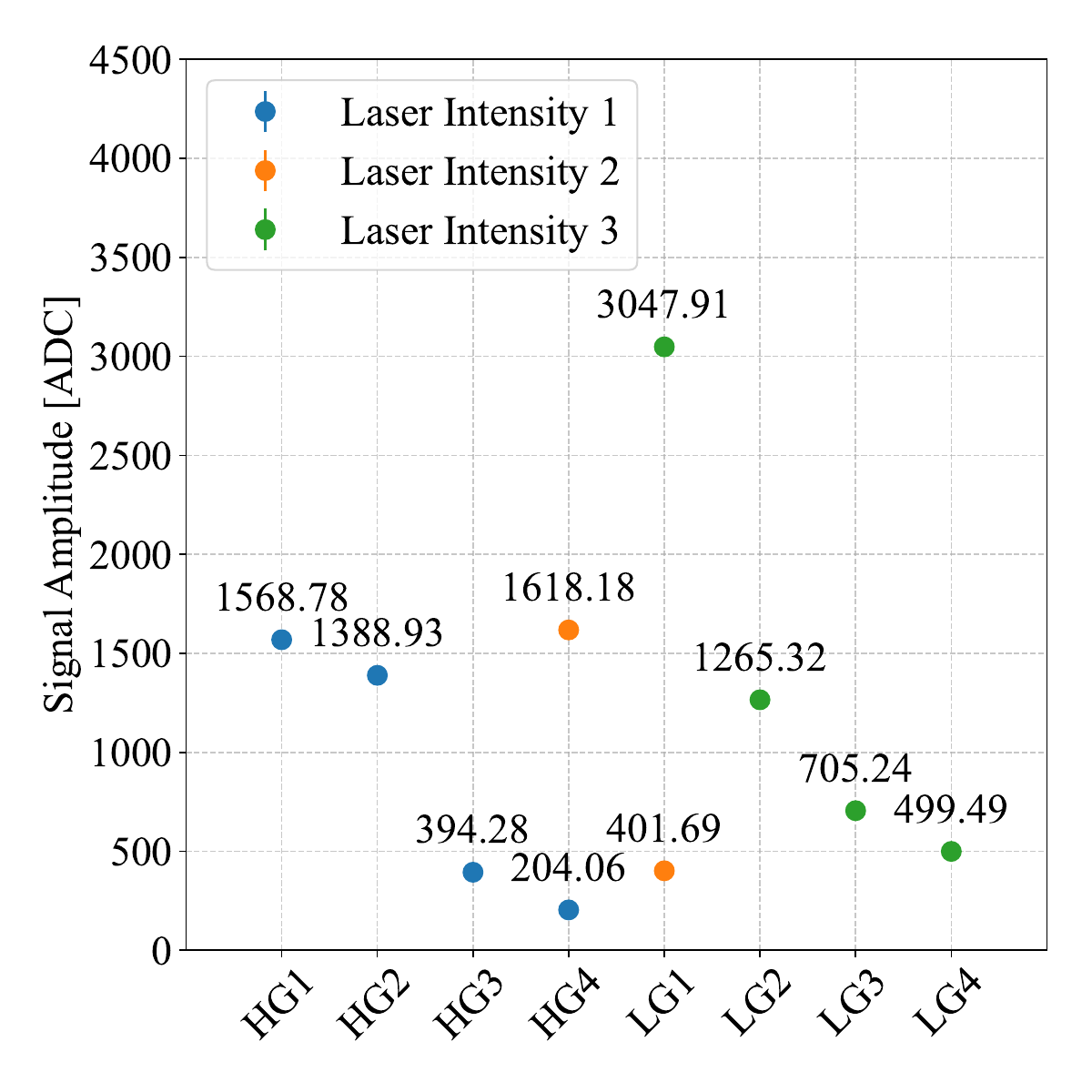}
\includegraphics[width=.48\textwidth]{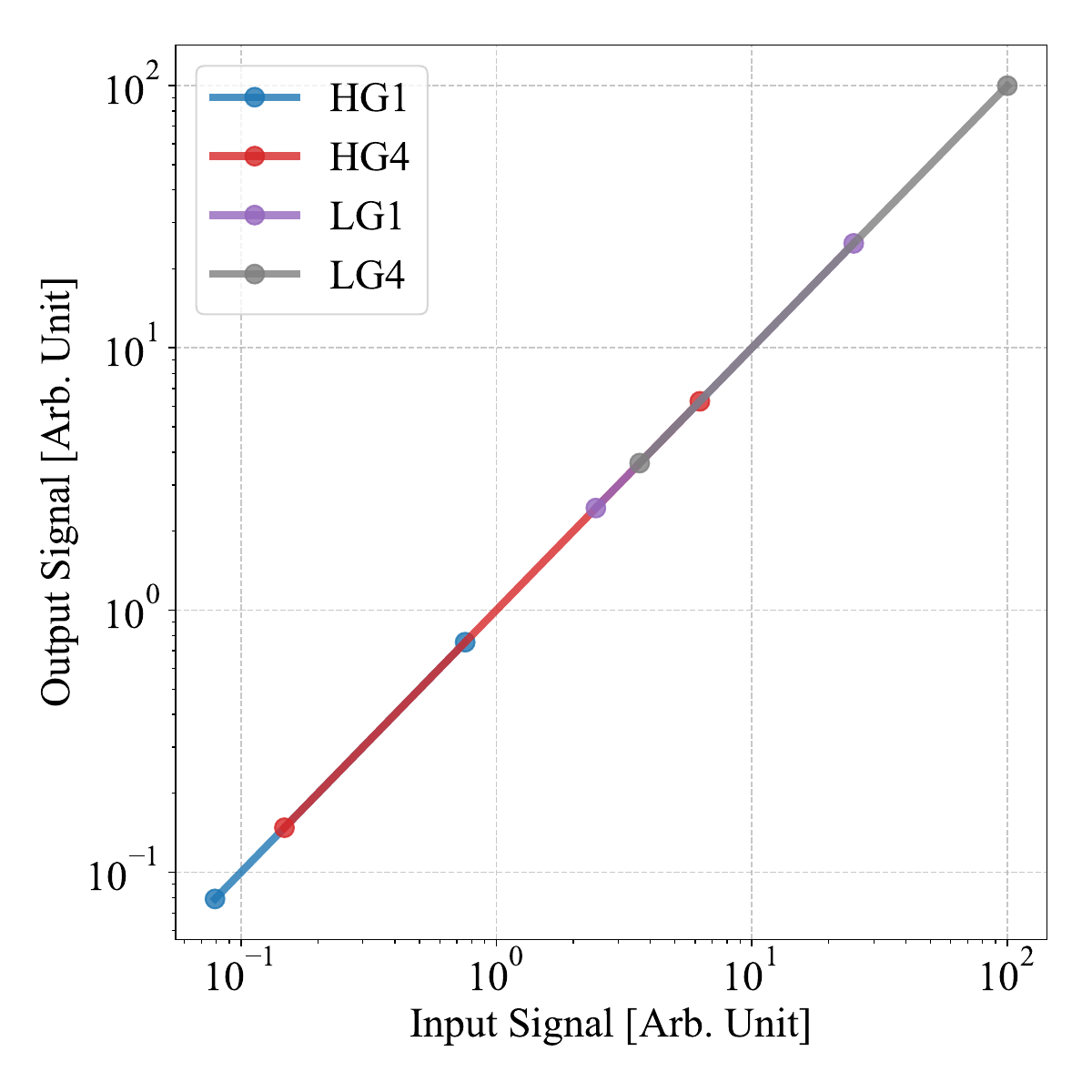}
\caption{Left: the chip's response to the laser signals collected by the SiPM under different gain modes. Right: the expected full dynamic range of the chip after inter-calibration.\label{fig:LaserInterCali}}
\end{figure}

The left plot in Figure~\ref{fig:LaserInterCali} shows the chip’s response to the laser signals detected by the SiPM. The electronic pedestal has already been subtracted, and the ratio between the values corresponding to each point gives the inter-calibration factors. To translate the signal amplitude from the highest gain (HG Mode 1) to the lowest gain (LG Mode 4), a minimum of three inter-calibration cycles are required, yielding a scaling factor of 188.98. The right plot of Figure~\ref{fig:LaserInterCali} illustrates this calibration outcome, where the lines indicate the linear response ranges corresponding to each gain mode. This plot further validates that inter-calibration enables the linear regime to cover the full dynamic range of the chip.

The laser inter-calibration experiment successfully determined a ratio coefficient of 188.98 between the highest and lowest gains, conclusively quantifying the tunable gain range of the chip. The inter-calibration factors enable conversion of signal amplitudes across different gain settings.

\subsection{Noise Performance}
The noise performance of this chip is primarily characterised by single photon sensitivity and the equivalent charge of the electronic pedestal noise. Single photon spectrum measurements were performed using small signals at the level of a few \unit{\mathrm{p.e.}}, to evaluate the signal-to-noise ratio and single photon calibration capability of the detector system. The experimental setup utilised a Hamamatsu S13360-3025PE SiPM operating at a \qty{5}{\volt} over-voltage, integrated with an LED calibration system, while the chip operated under the highest gain setup (HG Mode 1). A pulsed LED emitting low-intensity light at around \qty{400}{\nano \meter} wavelength was driven by a signal generator at \qty{1}{\kilo \hertz}. The signal generator was synchronised with the chip through an external trigger to minimise random noise. Photon signals detected by the SiPM were then digitised by the chip. To characterise electronic pedestal noise, additional data acquisition was conducted with the LED turned off, recording the pedestal noise across all gain modes while the SiPM was connected.

\begin{figure}[tbp]
\centering
\includegraphics[width=.48\textwidth]{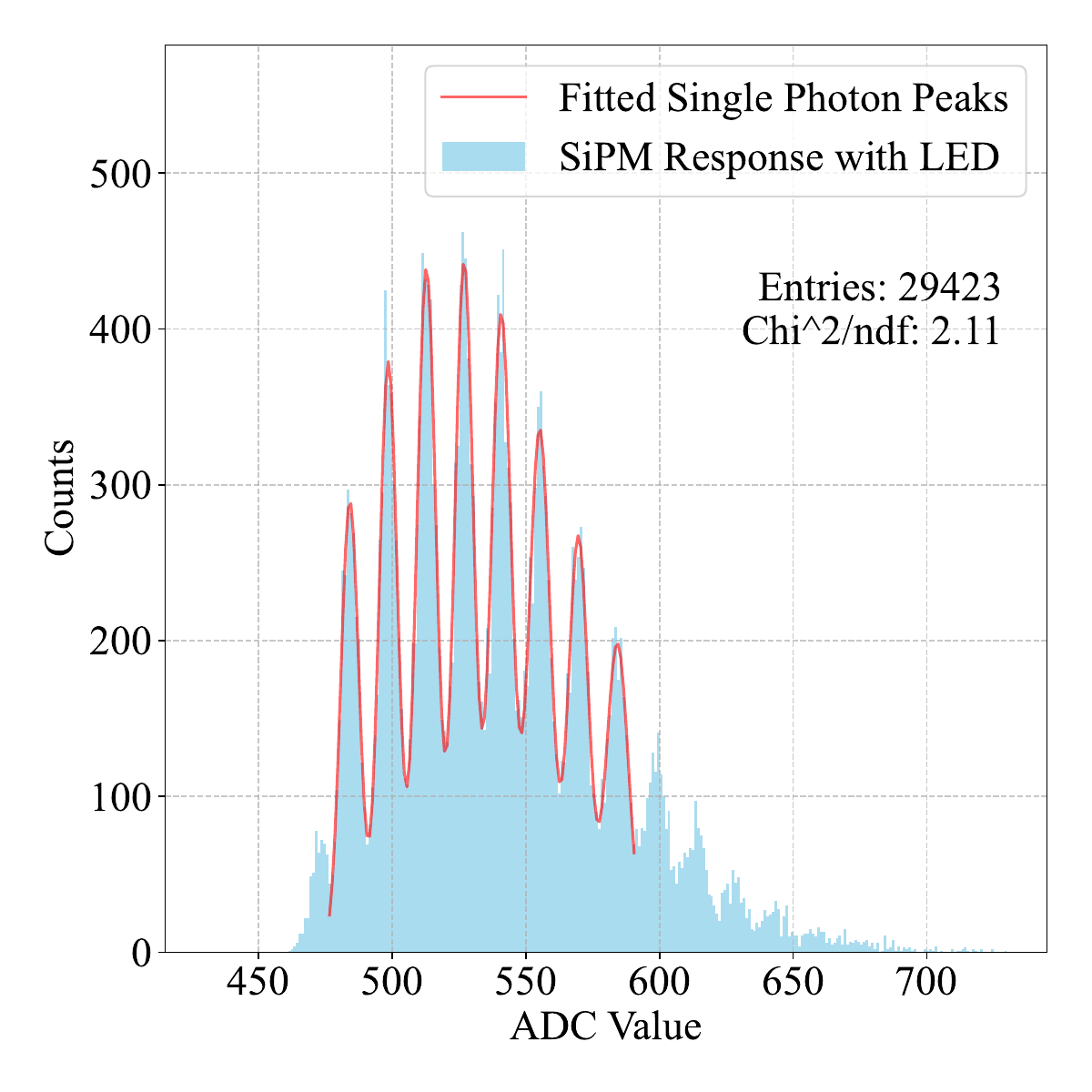}
\includegraphics[width=.48\textwidth]{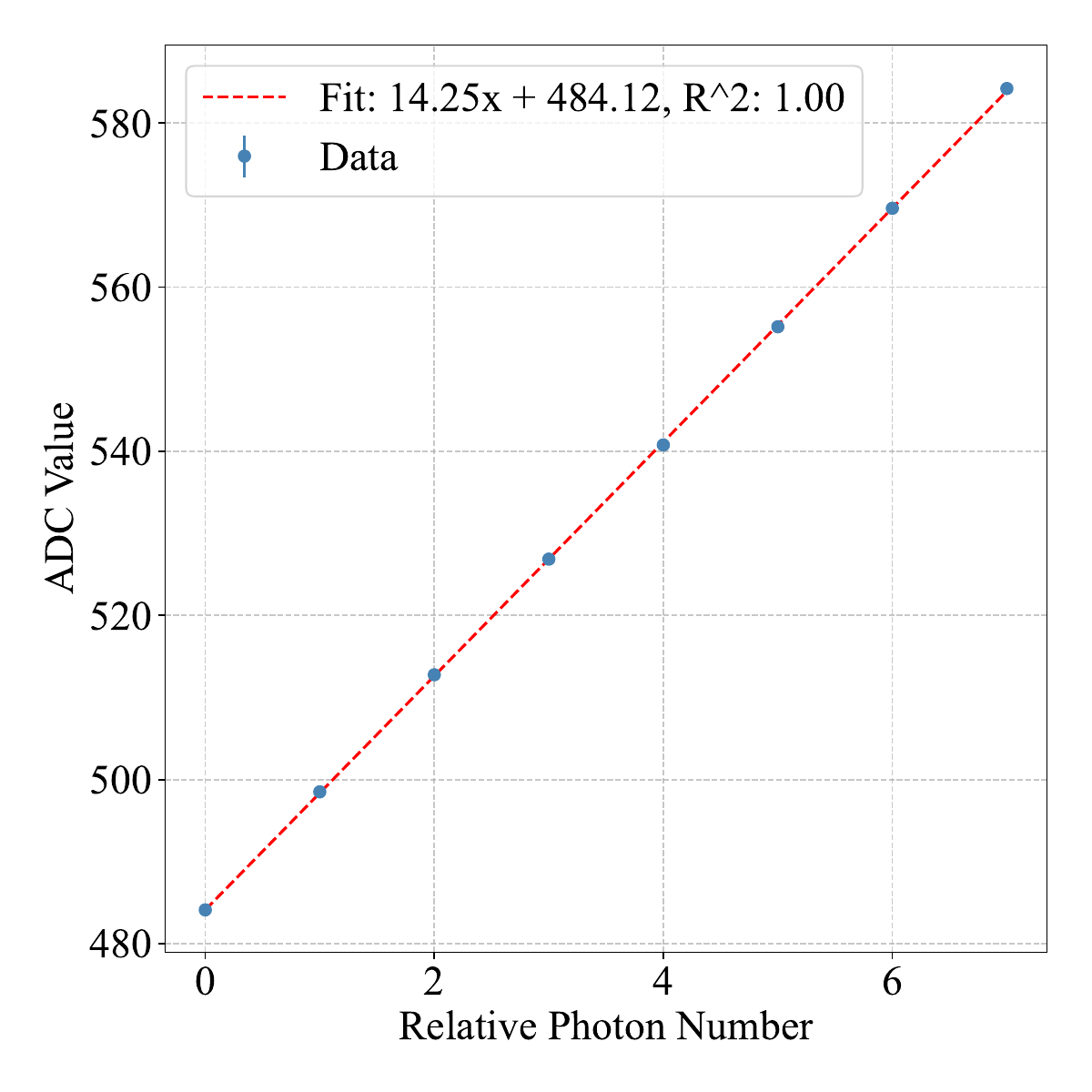}
\caption{Left: single photon spectrum of the Hamamatsu S13360-3025PE SiPM, measured with an LED source. Right: linear fit of the single photon peaks, where the slope corresponds to the single photon ADC value.\label{fig:SPSpectrum}}
\end{figure}

The measured single photon spectrum, shown in Figure~\ref{fig:SPSpectrum}, clearly resolves distinct photon peaks. Through multi-Gaussian and linear fitting, the signal amplitude corresponding to a single photon was determined to be \qty{14.25}{\mathrm{ADC}} counts. The signal-to-noise ratio in HG Mode 1 can be expressed as
\begin{equation}
    \text{SNR} = \frac{\langle \text{ADC}_\text{1p.e.} \rangle}{\sigma_{\text{ped}}}
\end{equation}
where $\langle \text{ADC}_\text{1p.e.} \rangle$ represents the average ADC counts corresponding to a single \unit{\mathrm{p.e.}}, and $\sigma_{\text{ped}}$ is the standard deviation of the Gaussian distribution of the electronic noise. Based on pedestal noise measurements, the SNR is calculated as 3.50. For the SiPMs with a nominal gain of \num{7e5} used in this study, a single \unit{\mathrm{p.e.}} corresponds to \qty{112}{\femto \coulomb} charge. With this information, the equivalent charge of the pedestal noise across all gain modes can be derived using the inter-calibration factors. The results are listed in Table~\ref{tab:PedestalNoise}.

The noise characterisation validated single photon detection capability with an SNR of 3.50. The results meet the requirements of the CEPC crystal ECAL. The electronic noise equivalent charge was quantified within \qty{32}{\femto \coulomb} to \qty{1.7}{\pico \coulomb}.

\begin{table}[tbp]
\centering
\caption{Electronic pedestal noise performance of MPT2321 with a typical SiPM connected.\label{tab:PedestalNoise}}
\smallskip
\begin{tabular}{c|c|c}
\hline
\textbf{Configuration} & \textbf{Pedestal Std. Dev. [\unit{\mathrm{ADC}}]} & \textbf{Noise Equivalent Charge [\si{\femto \coulomb}]}\\
\hline
HG Mode 1 & 4.08 & 32.07\\
HG Mode 2 & 3.70 & 32.85\\
HG Mode 3 & 1.72 & 53.79\\
HG Mode 4 & 1.51 & 91.24\\
\hline
LG Mode 1 & 1.13 & 275.06\\
LG Mode 2 & 1.12 & 668.42\\
LG Mode 3 & 1.11 & 1167.70\\
LG Mode 4 & 1.15 & 1738.61\\
\hline
\end{tabular}
\end{table}

\section{Beamtest for Dynamic Range Studies}
\label{sec:BeamTests}
To evaluate the dynamic range of the chip, and thereby assess its application potential in the high-granularity crystal ECAL, a beamtest was carried out at the Deutsches Elektronen-Synchrotron (DESY) TB22 beamline~\cite{cite:DESYTB}.

\subsection{Experimental Setup}
\begin{figure}[htbp]
\centering
\includegraphics[width=.48\textwidth]{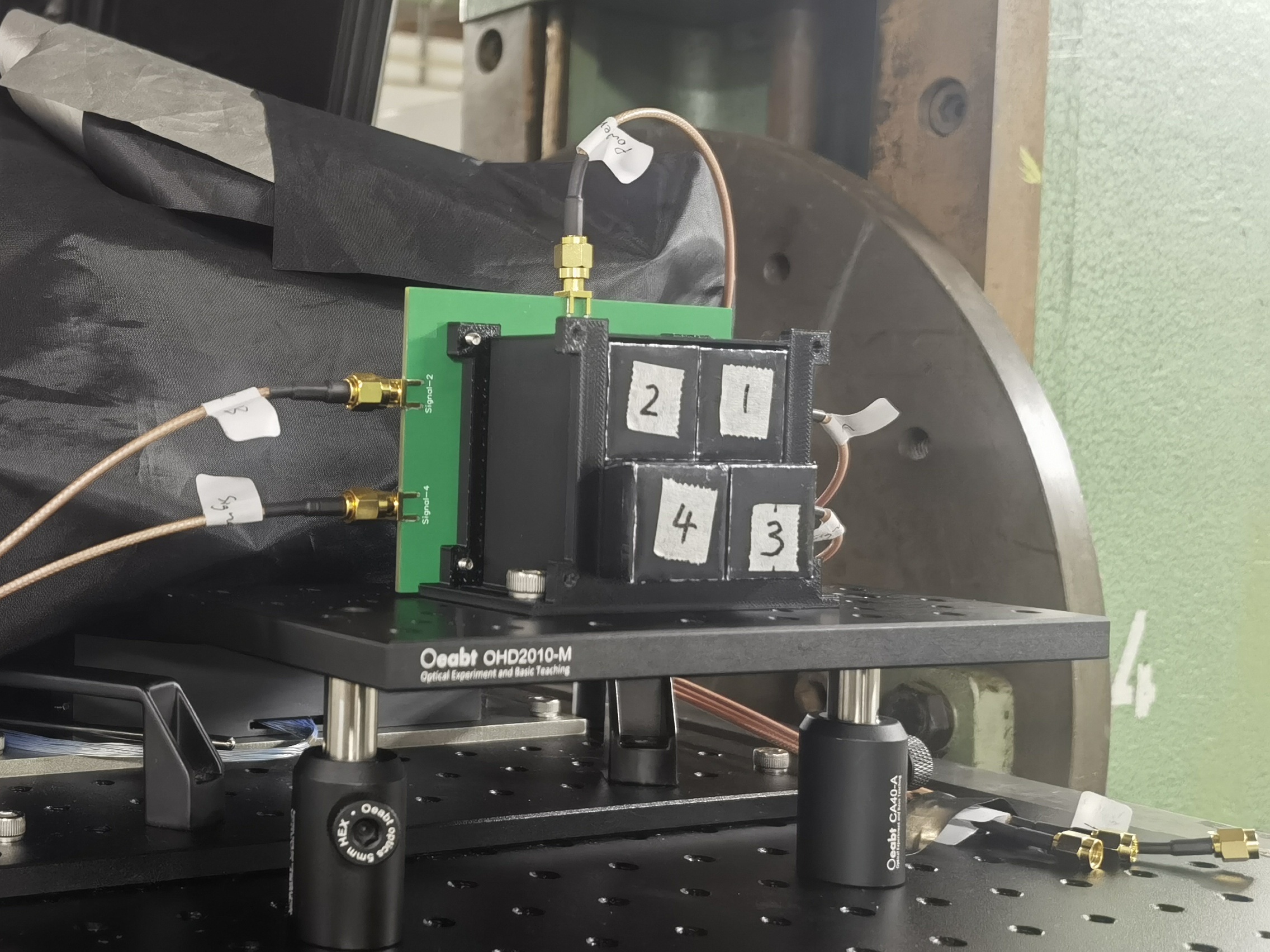}
\includegraphics[width=.48\textwidth]{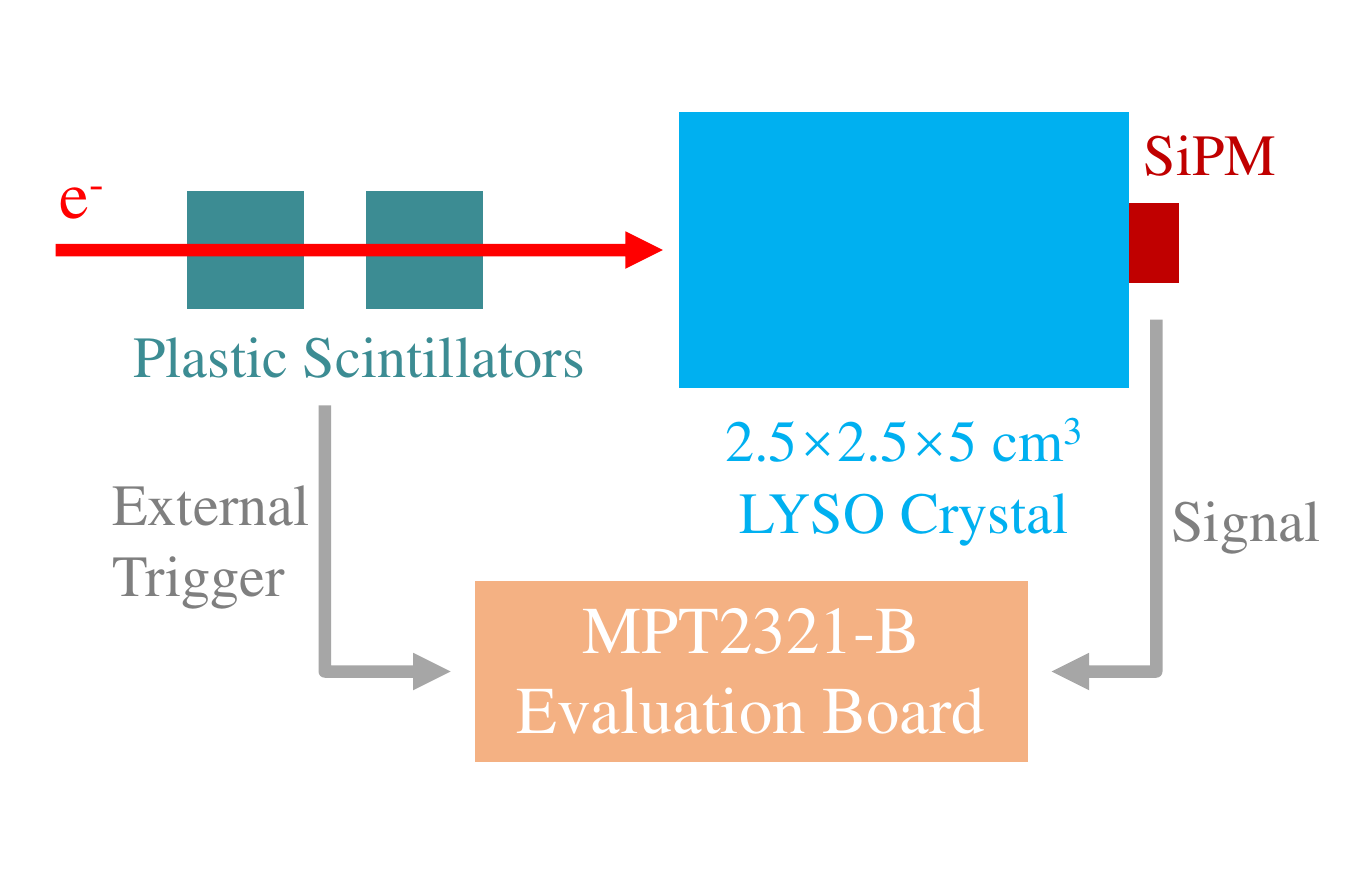}
\caption{Left: a photo of the LYSO crystal on the test platform for dynamic range studies, with the readout SiPM attached on the rear surface. Right: schematic of the beamtest setup at DESY TB22.}
\label{fig:BeamtestSetup}
\end{figure}

The experiment employed Lutetium Yttrium Oxyorthosilicate (LYSO) crystals with dimensions of \qtyproduct{2.5 x 2.5 x 4}{\centi \meter \cubed} and \qtyproduct{2.5 x 2.5 x 5}{\centi \meter \cubed}. The LYSO crystals exhibit a high light yield of more than \qty{30000}{\ensuremath{\gamma} \per \mega \electronvolt}, ensuring a sufficient light output for the dynamic range evaluation. The crystals were wrapped in Enhanced Specular Reflector (ESR) film and coated with aluminium foil to maximise light collection, before being air-coupled to the Hamamatsu S13360-3025PE SiPMs. The SiPMs were operated at a nominal over-voltage of \qty{5}{\volt} and under a room temperature of approximately \qty{23}{\celsius}. The setup of the beamtest is shown in Figure~\ref{fig:BeamtestSetup}. 3D-printed holders were used to provide support for the crystals and the PCBs. Upstream of the crystals, a coincidence trigger system, consisting of two \qty{1}{\centi \meter \cubed} plastic scintillator cubes, was used to collimate the beam and provide an external trigger signa. \qty{5}{\giga \electronvolt \per \mathit{c}} electrons were incident along the length of the crystals. Data acquisition was conducted under LG Mode 4, which offers the lowest gain of the chip. Additionally, electronic pedestal and single photon calibrations were performed at the TB22 site to ensure all calibration data were acquired under nearly constant environment temperature.

\subsection{Data Analysis}
With \qty{5}{\giga \electronvolt \per \mathit{c}} electron beams, the light signals generated by the crystal were collected by the SiPMs and then digitised by the chip. During the experiments, it was observed that the response of the \qty{4}{\centi \meter} LYSO crystals was insufficient for assessing the maximum dynamic range of the chip. Thus, the beamtest primarily focuses on the results corresponding to the \qty{5}{\centi \meter} crystals. 

A series of calibrations were performed on the raw data to quantify the dynamic range. The electronic pedestal was subtracted from the raw data, using the events acquired from random triggering. With the single photon calibration data taken under HG Mode 1 and the inter-calibration factor between the highest gain and the lowest gain mentioned in Section~\ref{sec:GainInterCali}, the single photon ADC value under LG Mode 4 was obtained. The ADC data from the beamtest can thus be converted into the number of \unit{\mathrm{p.e.}} using this calibration factor. The relationship between the crystal scintillation time constant and the ASIC response will not be discussed here, as the decay time of the LYSO crystal is typically around \qty{40}{\nano \second} and generally does not have a significant impact.

Figure~\ref{fig:BeamtestNumberPhoton} shows the detector’s response to the electron beam after the calibrations. Considering the following limiting factors present in the experiment: (a) substantial energy leakage with a limited-size crystal, (b) non-linear response of the readout chip, (c) saturation effect of the SiPM, and (d) momentum spread of the electron beam, this distribution is unsuitable as a reference for the detector's energy measurement performance. The result is only intended for the investigation of the dynamic range with the peak value. By simply employing the Crystal Ball function~\cite{cite:CrystalBallFunc} to fit this distribution, the photon count and ADC value corresponding to the peak can be derived. Thus, a comprehensive analysis of the dynamic range can be conducted using the response curve presented in Section~\ref{sec:ResponseLinearity}.

\subsection{Results and Discussions}
\begin{figure}[htbp]
\centering
\includegraphics[width=.8\textwidth]{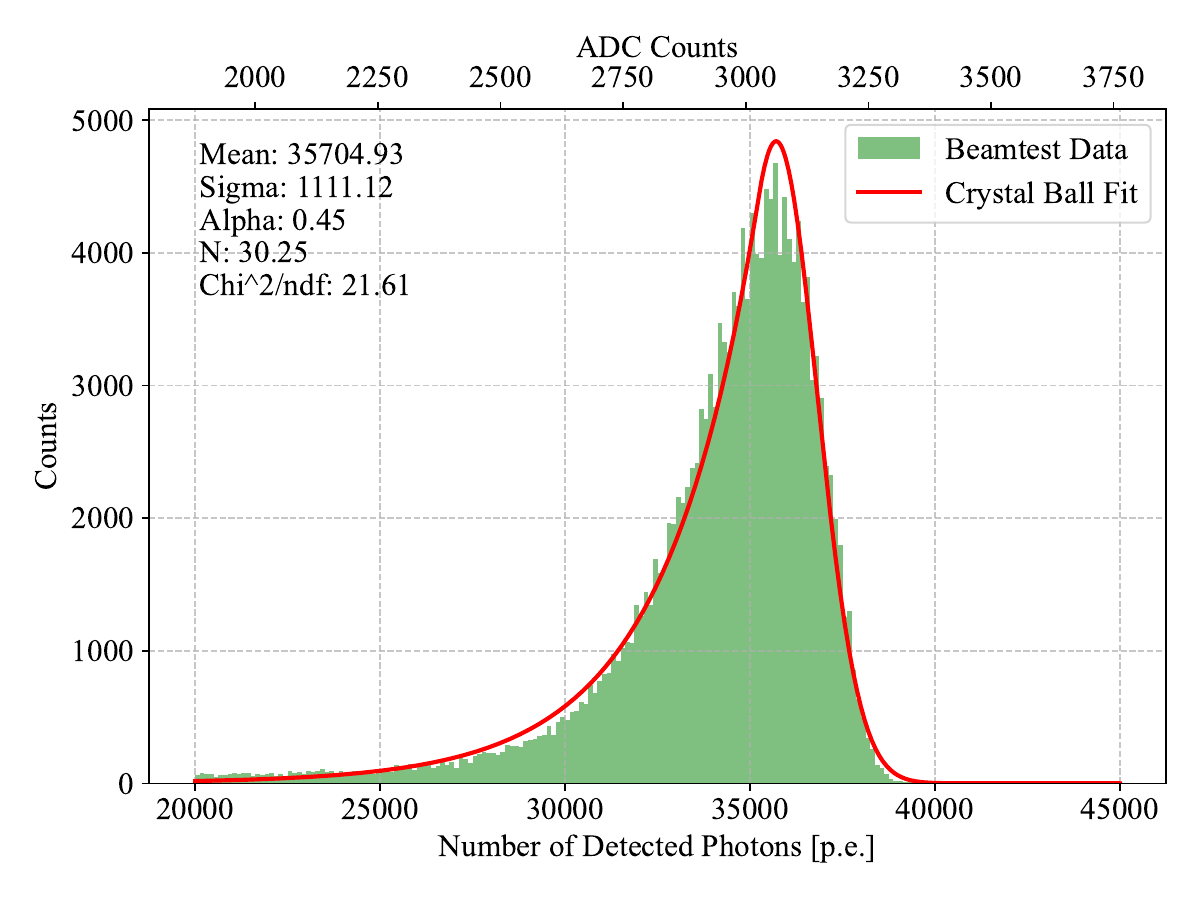}
\caption{Response of the \qty{5}{\centi \meter} LYSO crystal to \qty{5}{\giga \electronvolt \per \mathrm{c}} electrons. The fitting was performed using the Crystal Ball function with the unbinned likelihood method.}
\label{fig:BeamtestNumberPhoton}
\end{figure}

The peak of the distribution in Figure~\ref{fig:BeamtestNumberPhoton} corresponds to an ADC value of approximately \num{3050} and a photon count of \qty{35705}{\mathrm{p.e.}} This signal is equivalent to an injected charge value of nearly \qty{2000}{\pico \coulomb} according to Figure~\ref{fig:ChargeInjection}. Referencing against the linear ranges listed in Table~\ref{tab:LinearResponseResults}, the result has already entered the non-linear response region of the chip. Since the maximum linear range of the LG Mode 4 extends to \qty{2615}{\mathrm{ADC}}, the equivalent photon count within the linear response range can be calculated as approximately \qty{29936}{\mathrm{p.e.}} In such cases, considering the nominal SiPM gain of \num{7e5} (\qty{112}{\femto \coulomb} per \unit{\mathrm{p.e.}}), the total charge collected by the chip reaches \qty{3.35}{\nano \coulomb}.

It is apparent that the number of detected photons obtained from the beamtest significantly exceeds the nominal pixel number (\num{14400}) of the SiPM. This non-linear response behaviour is related to the intrinsic properties of the crystal and the SiPM. While the saturation effect of the SiPM limits the number of pixels fired, the long-tail temporal distribution of the scintillation photons from the LYSO crystal allows for the SiPM pixels to be retriggered. Thus, the measured response of the SiPM can be significantly increased. A comprehensive study of this phenomenon has been conducted through Monte Carlo simulations, as detailed in Appendix~\ref{sec:MCSimulation}.

\begin{figure}[tbp]
\centering
\includegraphics[width=.8\textwidth]{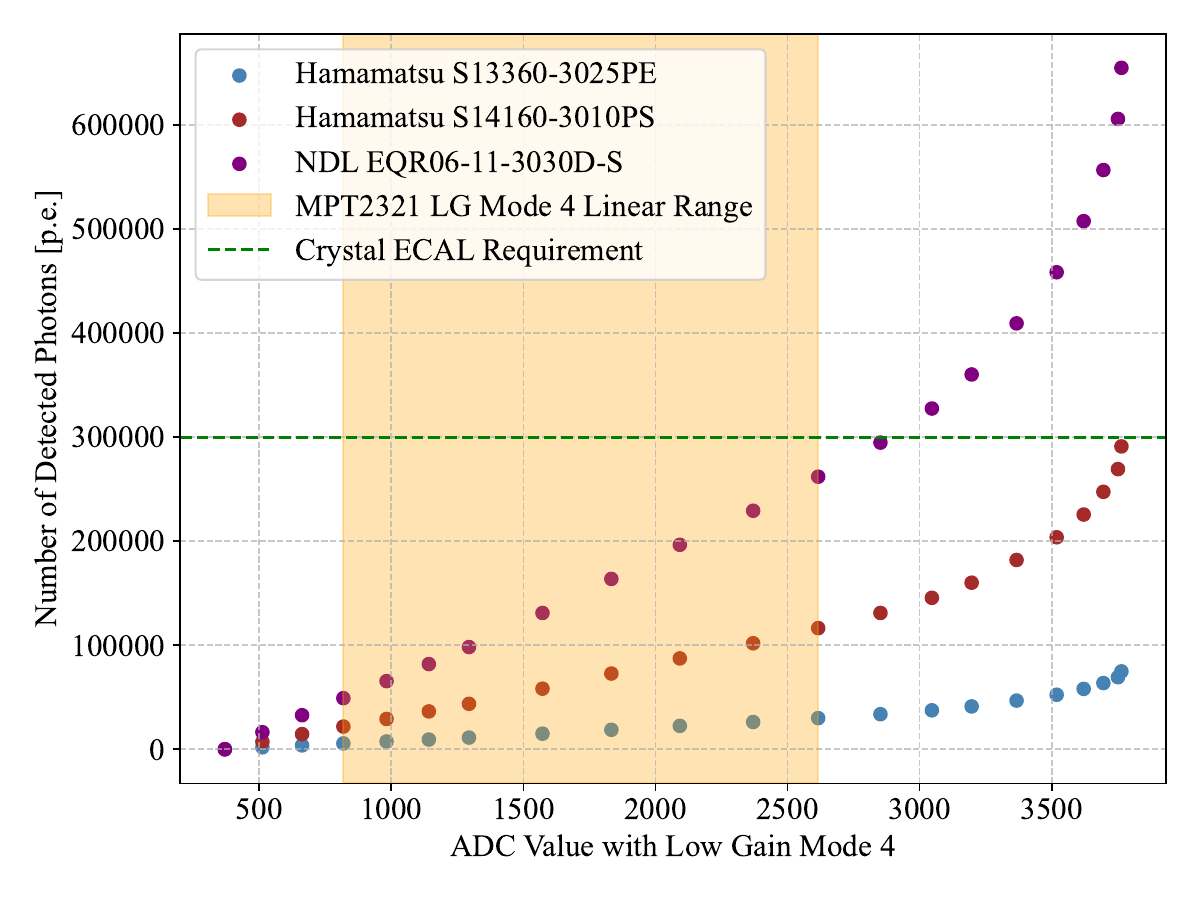}
\caption{The relationship between the ADC values and the corresponding number of detected photons for different SiPM models with the MPT2321 chip. The orange region indicates the linear response area with LG Mode 4, while the green line represents the dynamic range requirement for the CEPC crystal ECAL.}
\label{fig:ADCvsPE}
\end{figure}

Since the gain directly determines the amount of charge output to the chip, using the SiPM with a lower gain can effectively extend the dynamic range. In this case, achieving a good SNR will be a major technical challenge since a relatively high gain is required for single photon calibrations. If the feasibility of single photon calibration is not considered yet, the chip's dynamic range with different SiPM models can be predicted by linear extrapolation using the SiPM gain values. Figure~\ref{fig:ADCvsPE} shows the achievable dynamic range with different SiPMs. It is derived by scaling the chip's response curves from ADC counts to \unit{\mathrm{p.e.}}, and considering the gain values of different SiPMs. Since the ASIC focuses on the charge output from the SiPM, the non-linear effect of the SiPM does not contribute to the conclusions. Therefore, only the non-linearity of the electronics is considered.

As shown in Figure~\ref{fig:ADCvsPE}, the Hamamatsu S13360-3025PE SiPM (\num{14400} pixels, \num{7e5} nominal gain) is the device employed in this beamtest, demonstrating a capability of detection of about \qty{2.99e4}{\mathrm{p.e.}} within the linear range of the chip. Employing Hamamatsu S14160-3010PS~\cite{cite:HPKS14160} SiPM (\num{89984} pixels, \num{1.8e5} nominal gain) will extend the dynamic range to over \qty{1.16e5}{\mathrm{p.e.}}, while collecting of over \qty{2.62e5}{\mathrm{p.e.}} can be achieved with NDL EQR06 11-3030D-S~\cite{cite:NDLEQR06} SiPM (\num{2447419} pixels, \num{8e4} nominal gain). The green line in Figure~\ref{fig:ADCvsPE} represents the dynamic range requirements for the crystal ECAL, specifically the maximum detectable number of photons. The result indicates that the chip can not fully meet the dynamic range requirement of the CEPC crystal ECAL with these SiPM models. Nevertheless, since LG Mode 4 of the chip still features a wide non-linear range before full saturation, the measurement of larger signals with the non-linear part remains possible.

In general, the beamtest measurements reveal three critical insights: (a) The MPT2321 ASIC achieves a linear dynamic range of up to \qty{29936}{p.e.} (\qty{3.35}{\nano\coulomb}), when accounting for the scintillation decay time and SiPM pixel retriggering effects; (b) Analysis confirms that with low-gain SiPMs (\num{8e4}), the ASIC covers $87\%$ of the dynamic range required for the CEPC crystal ECAL; (c) The observed non-linear regime (>\qty{2615}{\mathrm{ADC}}) remains usable for energy reconstruction if properly calibrated. These results emphasise the critical role of synergistic optimisation between crystals, SiPMs, and readout systems in addressing dynamic range challenges.

\section{Summary and Outlook}
\label{sec:Summary}
A systematic investigation of the MPT2321 ASIC was conducted through laboratory tests and beam experiments to evaluate its performance for SiPM-based calorimetry. Response linearity tests across all gain modes confirmed operational ranges with deviations limited to $\pm5\%$. Inter-calibration studies using SiPM laser tests established scaling factors among different gain modes. Noise characterisation demonstrated the single photon detection capability of the chip, with a signal-to-noise ratio of 3.50 in the highest gain mode. The noise equivalent charge has been quantified for all gain modes.

Beam experiments validated the ASIC's promising charge detection capability up to \qty{3.35}{\nano \coulomb} (\qty{29936}{\mathrm{p.e.}}, \num{7e5} SiPM gain) within the linear region. However, the current design fails to meet the dynamic range requirements of the CEPC crystal ECAL, even when employing a SiPM with a considerably low gain of \num{8e4}.

Future work will focus on implementing the ASIC in multi-channel configurations and conducting experiments with the crystal ECAL prototype to investigate its physics performance. Additionally, non-linear calibration method, not investigated in this study, requires further exploration to extend the ASIC's dynamic range for high-energy applications.

\appendix
\section{Simulation Studies for the SiPM Non-linear Response}
\label{sec:MCSimulation}
Saturation effects are expected to have a significant impact on the measurements with a limited number of SiPM pixels. Considering the scintillation decay time of the crystal and the pixel recovery time of the SiPM, the output signals of the crystal-SiPM units can exceed the SiPM pixel number. To enhance the understanding of the beamtest results, modelling and simulation of the SiPM response have been performed.

A Geant4~\cite{cite:Geant4} simulation tool was developed with the implementation of a single \qtyproduct{2.5 x 2.5 x 5}{\centi \meter \cubed} LYSO crystal. \qty{5}{\giga \electronvolt \per \mathit{c}} electrons give an energy deposition of roughly \qty{900}{\mega \electronvolt} in the crystal. An optical simulation was conducted, fully incorporating the crystal's scintillation properties, optical parameters, surface conditions, and wrapping materials. The timestamps and wavelengths of each optical photon hit on the SiPM were recorded for further investigations.

Following the crystal optical simulation, comprehensive modelling of the Hamamatsu S13360-3025PE SiPM response was developed~\cite{cite:SiPMSimulation}. Different from modelling the entire waveform output of SiPM, the toy Monte Carlo study aims to provide a simplified model of the SiPM responses to photon hits with specific time distributions. This model incorporates several critical steps. Based on the results of the optical simulation, photons can be sampled with specific timestamps and wavelengths. Then, taking into account the geometry and the wavelength-dependent photon detection efficiency (PDE) of the SiPM, whether a specific pixel will be fired can be determined. Upon the detection of a photon, the model assesses the recovery status of the pixel and further evaluates the potential crosstalk effect. The contribution of dark noise has not been included. Finally, the non-linear response of the crystal-SiPM units can be obtained.

\begin{figure}[tbp]
\centering
\includegraphics[width=.8\textwidth]{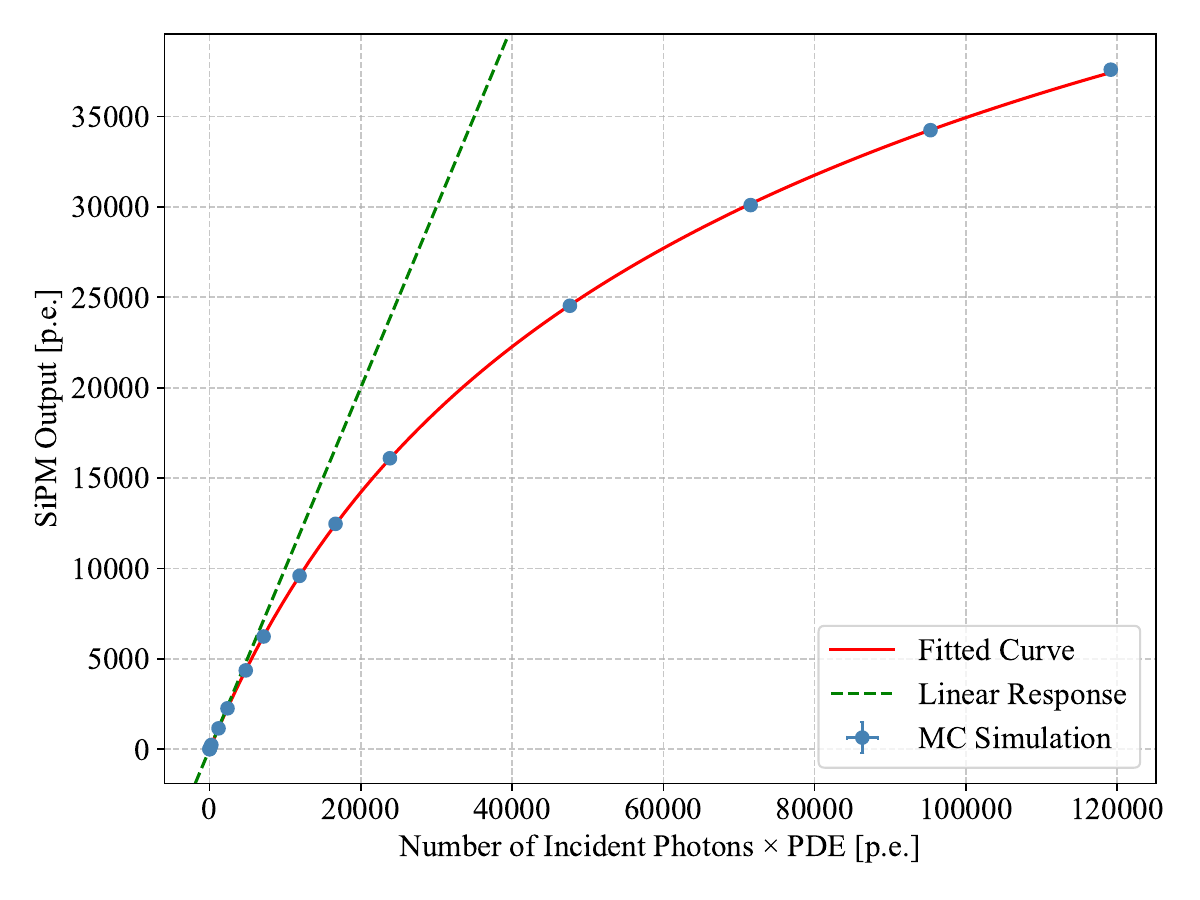}
\caption{Toy Monte Carlo simulation of the non-linear response of the crystal-SiPM units. The horizontal axis represents the expected number of photons detected by a SiPM without noise and the pixel recovery state, while the vertical axis represents the actual output signal of the SiPM.}
\label{fig:MCSiPMResponse}
\end{figure}

Figure~\ref{fig:MCSiPMResponse} illustrates the relationship between the expected number of photons detected by the SiPM that is free from any noise effect and has zero recovery time, and the realistic output generated by the SiPM and delivered to the front-end electronics. These points have been fitted with the function described in Reference~\cite{cite:SiPMFit}. Given a SiPM output of \qty{35705}{\mathrm{p.e.}}, there are approximately \num{105503} photons expected to be detected by an ideal SiPM without noise and saturation, considering only the PDE effect. This value is generally consistent with the result given by the Geant4 optical simulation.

\acknowledgments
The beam experiments have been performed at the Test Beam Facility at DESY, Hamburg, Germany. The authors would like to thank the DESY Test Beam Facility operation teams for their technical assistance, and the CALICE Collaboration~\cite{cite:CALICE} for their support in the beamtest experiments. Special acknowledgment is extended to Yan Huang, Wei Shen and other colleagues from MicroParity Tech. Inc. for their technical guidance on chip application.

\end{document}